\documentstyle[aps,prb,graphicx,multicol,epsfig]{revtex}

\def\bs#1{\bbox{#1}}
\def\ve#1{\bs{\mathbf{#1}}}
\newcommand{\be}{\begin{eqnarray}}
\newcommand{\ee}{\end{eqnarray}}

\begin{document}

\draft \title{Selective nanomanipulation using optical forces}

\author{Patrick C. Chaumet}

\address{Institut Fresnel (UMR 6133), Facult\'e des Sciences et
Techniques de St J\'er\^ome, Av. Escadrille Normandie-Niemen, F-13397
Marseille cedex 20}

\author{Adel Rahmani}

\address{Laboratoire d'Electronique, Opto\'electronique et 
Microsyst\`emes-UMR CNRS 5512-Ecole Centrale de Lyon 36, avenue Guy de 
Collongue, BP 163, F-69131 Ecully Cedex, France}

\author{Manuel Nieto-Vesperinas}

\address{Instituto de Ciencia de Materiales de Madrid, Consejo
Superior de Investigaciones Cientificas, Campus de Cantoblanco Madrid
28049, Spain}

\maketitle

\begin{abstract}

We present a detailed theoretical study of the recent proposal for
selective nanomanipulation of nanometric particles above a substrate
using near-field optical forces [Chaumet {\it et al.}
Phys. Rev. Lett. {\bf 88}, 123601 (2002)].  Evanescent light
scattering at the apex of an apertureless near-field probe is used to
create an optical trap. The position of the trap is controlled on a
nanometric scale via the probe and small objects can be selectively
trapped and manipulated.  We discuss the influence of the geometry of
the particles and the probe on the efficiency of the trap. We also
consider the influence of multiple scattering among the particles on
the substrate and its effect on the robustness of the trap.

\end{abstract}


\begin{multicols}{2}

\section{Introduction}

Thirty years ago, it was demonstrated by Ashkin that optical fields
produce a net force on neutral particles.~\cite{ashkin69,ashkin70}
Since then it has been shown that it was possible to exploit the
mechanical action of optical fields in a wide range of
applications. From atomic and nonlinear physics to biology, optical
forces have provided a convenient way to manipulate, non
destructively, small particles in a liquid
environment~\cite{ashkin86,ashkin87,block89,ashkin97}. These optical
forces can also be used to create microstructures by optical
binding,~\cite{burns} or measure the van der Waals force between a
dielectric wall and an atom.~\cite{landragin} But one of the most
interesting applications of optical forces is the the optical
tweezers. They have proved useful not only for trapping particles, but
also for assembling objects ranging from microspheres to biological
cells \cite{assembling,macdonald} (notice that in
Ref.~[\ref{macdonald}] the trapped spheres are 50 times larger that
the wavelength used in the experiment). More recently, optical
tweezers have been used to transport Bose-Einstein condensates over
large distance.\cite{gustavson} However, most of these manipulations
involve objects whose size is of the order of one to several
micrometers. While for much smaller objects, such as atoms or
molecules, the scanning tunneling microscope provides a powerful tool
for manipulation and engineering\cite{stm}, dealing with neutral
particles of a few nanometers requires new experimental approaches.

A novel approach was presented recently, where an apertureless
near-field probe is used to create localized optical traps and allow
for the selectively capture and manipulation of nanoparticles in
vacuum or air above a substrate .~\cite{prl} In this paper we analyze
in detail the scheme presented in Ref. \onlinecite{prl} and we discuss
the interplay of the different physical processes that contribute to
the force experienced by the particles (including van der Waals
forces). The particles are not in a liquid environment, hence there is
no Brownian motion (which would otherwise induce a disruptive force
for small particles) and the apertureless probe can be used as a
near-field optical probe to localize and select the
particles~\cite{zenhausern,defornel}.

In Sec. II we describe briefly the method used to compute the optical
forces. In Sec. III A we study the optical force experienced by a
sphere in presence of a tungsten tip. First we explain the principle
of the manipulation of a nano-object with the apertureless probe and
then we look at the influence of the different parameters of the
system, (geometry of the tip, size of the nanoparticle, illumination)
on the trapping.  In Sec. III B the presence of many particles on the
substrate is investigated to study the influence of neighbors on the
manipulation of a particle. Finally in Sec. IV we present our
conclusions. In appendix A we underline the importance of using total
internal reflection to get an efficient optical trap at the tip apex,
and in appendix B we compare the optical force with the other forces
present in this system (gravitational force, van der Waals force,
electrostatic force, and capillary force).

\section{Computation of the optical forces}

The theory used to compute the optical forces has been presented
previously.~\cite{chaumet1}. We use the couple dipole method
(CDM). Here we only recall the main steps. First, the coupled dipole
method~\cite{purcell,chaumet3} is used to derive the field inside the
different objects (probe and particles).  Each object is discretized
into dipolar subunits and the field at each subunit satisfies the
following self-consistent equation
%%%%%%%%%%%%%%%%%%%%%%%%%%%%%%%%%%%%%%%%%%%%%%%% 
\be \label{dipi}
\ve{E}(\ve{r}_i,\omega) & = & \ve{E}_0(\ve{r}_i,\omega) +
\sum_{j=1}^{N} [\ve{S}(\ve{r}_i,\ve{r}_j,\omega)\nonumber\\ & + &
\ve{T}(\ve{r}_i,\ve{r}_j,\omega)] \alpha_j(\omega)
\ve{E}(\ve{r}_j,\omega).\ee
%%%%%%%%%%%%%%%%%%%%%%%%%%%%%%%%%%%%%%%%%%%%%%%% 
$\alpha_j(\omega)$ is the dynamic polarizability of subunit
$j$,~\cite{draine}, $\ve{T}$ is the field linear response to a dipole
in free space,~\cite{jackson75,rahmani1} and $\ve{S}$ the field linear
response to a dipole, in the presence of a
substrate.~\cite{agarwal,rahmani2} Note that the field obtained in
Eq.(\ref{dipi}) takes into account all the multiple interactions
between the particles, the substrate, and the tip.  The second step is
to derive the optical forces experienced by each subunit.  Once the
electric field is known, the component of the total force~\cite{time}
on the $i$th subunit is given by
%%%%%%%%%%%%%%%%%%%%%%%%%%%%%%%%%%%%%%%%%%%%%%%%% 
\be \label{forcec}
F_u(\ve{r}_i)=(1/2)\Re e\left(\sum_{v=1}^{3} p_v(\ve{r}_i,\omega)
\frac{\partial (E^v(\ve{r}_i,\omega))^*}{\partial u}\right), \ee
%%%%%%%%%%%%%%%%%%%%%%%%%%%%%%%%%%%%%%%%%%%%%%%%% 
where $u$ or $v$, stand for either $x,y$, or $z$, and
$\ve{p}(\ve{r}_i,\omega)$ is the electric dipole moment of the $i$th
subunit.~\cite{chaumet2} Notice that the derivative of the field is
obtained by differentiating Eq.(\ref{dipi}). To compute the force
exerted by the light on any given object, one has to sum the force
experienced by each dipole forming the object. The main advantage of
using the CDM to compute the optical force is that retardation and
multiple scattering between the objects, the tip and the substrate are
accounted for.

\section{Results}

We consider a particle in glass, placed either in air or vacuum, with
relative permittivity $\varepsilon=2.25$ and a radius $a$, above a
dielectric substrate. The particle is illuminated by two evanescent
waves created by total internal reflection at the substrate/air
interface (angle of incidence $\theta>\theta_c=41.8^{\circ}$ with
$\sqrt{\epsilon}\sin\theta_c=1$ where $\epsilon=2.25$ is the relative
permittivity of the substrate). The importance of illuminating the
particle on the substrate with evanescent waves is explained in
appendix~\ref{A1}. The two evanescent waves are counterpropagating ,
i.e., $\ve{k}_{\parallel}=-\ve{k}'_{\parallel}$, with the same
polarization and a random phase relation (Fig.1). As discussed later,
this is to ensure a symmetric lateral force.  The optical trap is
created by the interaction of the incident waves with a tungsten probe
with a radius of curvature at the apex $r$.

\begin{figure}[ntb]
\begin{center}
\includegraphics*[draft=false,width=80mm]{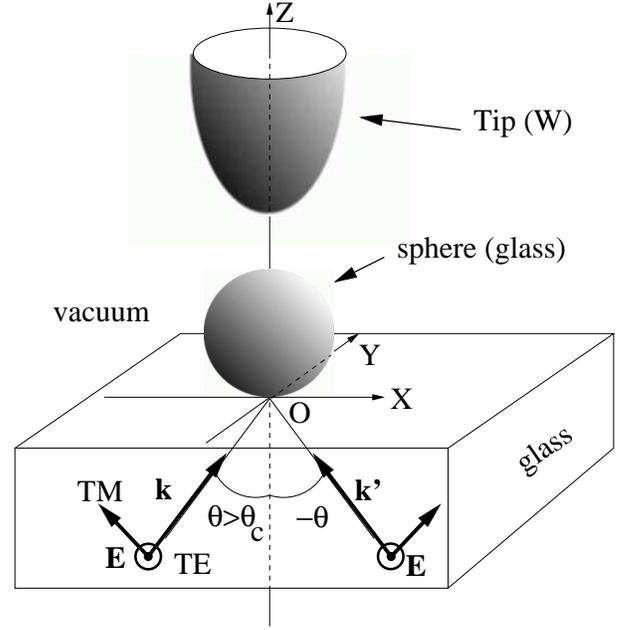}
\end{center}
\caption{Scheme of the configuration. A dielectric sphere (radius
10~nm) on a flat dielectric surface is illuminated under total
internal reflection. A tungsten probe is used to create an optical
trap.}
\end{figure}

Notice that all forces are computed for an irradiance of
0.05~W/$\mu$m$^2$, which correspond, for a laser with a power of 5~W,
to a beam focused over an area of 100~$\mu$m$^2$.

\subsection{isolated particle}

\subsubsection{principle of the manipulation}

In order to foster understanding of the selective trapping scheme, we
start by studying the interaction between a single sphere with radius
$a=10$~nm, and a tungsten tip ( a tip often used in apertureless
microscopy as they are not expensive and easy to prepare) with a
radius at the apex $r=$10~nm, which is a typical size for tips used in
experiments. The illumination wavelength is $\lambda=514.5$~nm. Figure
2 shows the $z$ component of the force experienced by the sphere
versus the vertical position of the tip above the sphere, for both TE
and TM polarizations. The illumination angle is close to the critical
angle, $\theta=43^{\circ}$.  As the tip gets closer to the sphere, the
evolution of the force is radically different for the two
polarizations.  The sphere experiences mainly three gradient forces
(because the sphere is small, the scattering force is negligible, and
since the relative permittivity is real, absorption does not
contribute to the force). The three forces are: first, the negative
gradient force due to the evanescent incident field (notice that for a
dielectric sphere, the gradient force always pushes the sphere toward
the region of high field intensity, as the evanescent field decays in
the direction of $z$ positive, the gradient force is negative),
second, the negative gradient force due to the interaction of the
sphere with itself via the substrate (this force can be understood as
the interaction between the dipole associated to the sphere and the
field at the dipole location, radiated by this dipole and reflected by
the surface; this force is always negative whatever the dielectric
constant of the sphere\cite{chaumet1}), and third, the gradient force
resulting from the interaction between the probe and the sphere. This
last gradient force can be either positive or negative.  For TM
illumination, there is a large enhancement of the field near the apex
of the probe due to the discontinuity across the air/tungsten boundary
\cite{novotny}.  This enhancement generates a positive gradient force
which, at short distances, counterbalances the two negative
contributions (due to the interaction of the particle with itself via
the substrate, and the incident evanescent waves).  The inset in
Fig. 2a shows that the force experienced by the sphere changes sign
when the tip is 25~nm away from the sphere.  On the other hand, for TE
polarization (Fig.~2b), as the tip gets closer to the particle, the
magnitude of the $z$ component of the force increases while the force
remains negative (directed toward the substrate and away from the
tip), hence preventing any trapping. This due to a decrease of the
field at the tip apex for this polarization (one can see the
electromagnetic field around a gold tip apex in Ref.[\ref{novotny}]),
thus giving a third negative contribution to the gradient force.
Because the apex of the tip and the sphere are small compared to the
wavelength, the nature of the interaction between the tip and the
sphere can be understood by considering the tip and the sphere as two
dipoles.  In TM polarization, these two dipoles have two components,
parallel and perpendicular to the substrate. As shown in
Ref.[\ref{chaumet4}] two aligned dipoles tend to attract each other
and two parallel dipoles tend to repel each other. For the same
magnitude of the two components (parallel and perpendicular) the
attractive force due to the component perpendicular to the substrate
is twice that of the repulsive force due to the parallel
component. Hence in TM polarization the sphere is attracted by the
tip.  For the TE polarization however, the two dipoles are essentially
parallel to the substrate and the sphere experiences a negative
gradient force.  Notice that if we only use a single laser beam, a
lateral force would appear as shown in Fig.~3. For TE polarization,
the lateral force is very small (in the fN range) and negative showing
that it is mainly due to the gradient force arising from the presence
of the tip (the radiation pressure from the incident field always
gives a force in the direction of the wave vector, hence in this case
a positive force). When the sphere is in contact with the substrate,
the lateral force ($x$ component) is weaker than the $z$ component of
the force by a factor 40.  As the static friction coefficient is one
(glass on glass), the sphere cannot slide along the surface. Indeed,
as shown by Kawata and Sugiura,~\cite{kawata} for the sphere to slide
on the substrate, its radius has to be large enough for radiation
pressure to overcome the gradient force. For TM polarization the
lateral force may have a disruptive effect as it tends to push the
sphere away from the tip, particularly when $F_z$ becomes positive. In
order to avoid this problem, we introduce a second,
counterpropagating, evanescent wave with a random phase relation with
respect to the first wave. In this way, the sphere experiences no
lateral force when it is right underneath the tip.  Note that due to
the coherence time of the laser (e.g. 200~ps for an Argon laser), one
can suspect that the sphere experiences spatial fluctuations. We
compute these spatial fluctuations for a glass sphere with a radius
$a=10$~nm, trapped by an optical force $|\ve{F}|=4$~pN. From the
second law of Newton the distance covered by the sphere, during the
time of coherence, is equal to $\frac{|\ve{F}| t^2}{2m}=8$~pm.
Therefore, in any realistic configuration, the trapped particle will
only be sensitive to the time-averaged trapping potential, without
actually being perturbed by the laser fluctuations. If the sphere is
larger, its sensitivity to the spatial fluctuation becomes even
smaller due to its larger weight.

\begin{figure}[ntb]
\begin{center}
\includegraphics*[draft=false,width=75mm]{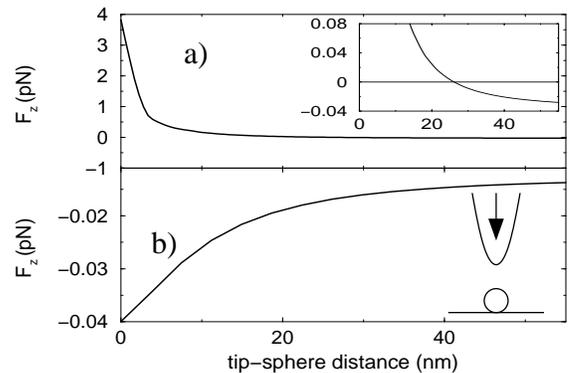}
\end{center}
\caption{$z$ component of the force experienced by the sphere versus
the distance between the tip and the sphere. a) TM polarization. The
inset is an enhancement of Fig. 2a near the sign reversal.  b) TE
polarization. The arrow indicated the direction along which the tip is
moved.}

\end{figure}
\begin{figure}[ntb]
\begin{center}
\includegraphics*[draft=false,width=70mm]{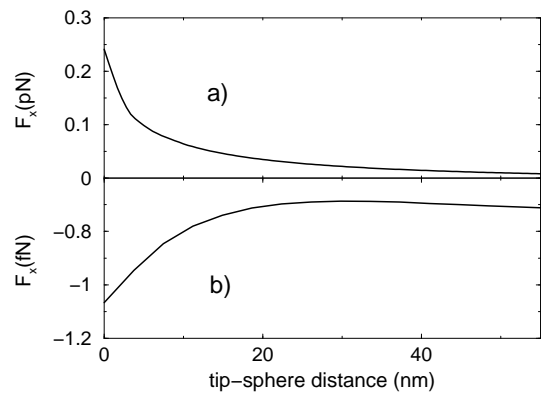}
\end{center}
\caption{$x$ component of the force if the symmetric illumination is
not used.  a) TM polarization.  b) TE polarization.}
\end{figure}

To assess fully the probe-particle coupling we need to study the
evolution of the force experienced by the particle, as the probe is
moved laterally. The coordinates $(x,y)$ represent the lateral
position of the sphere. The tip is at (0,0) (see Fig.~4a). Figure 4
shows the $z$ component of the force when the tip is 25~nm above
the substrate, for TE and TM polarizations, and for an angle of
incidence $\theta=43^{\circ}$. For TM polarization, Fig.~4b represents
the magnitude of the $z$ component of the force.

\begin{figure}[ntb]
\begin{center}
\includegraphics*[draft=false,width=80mm]{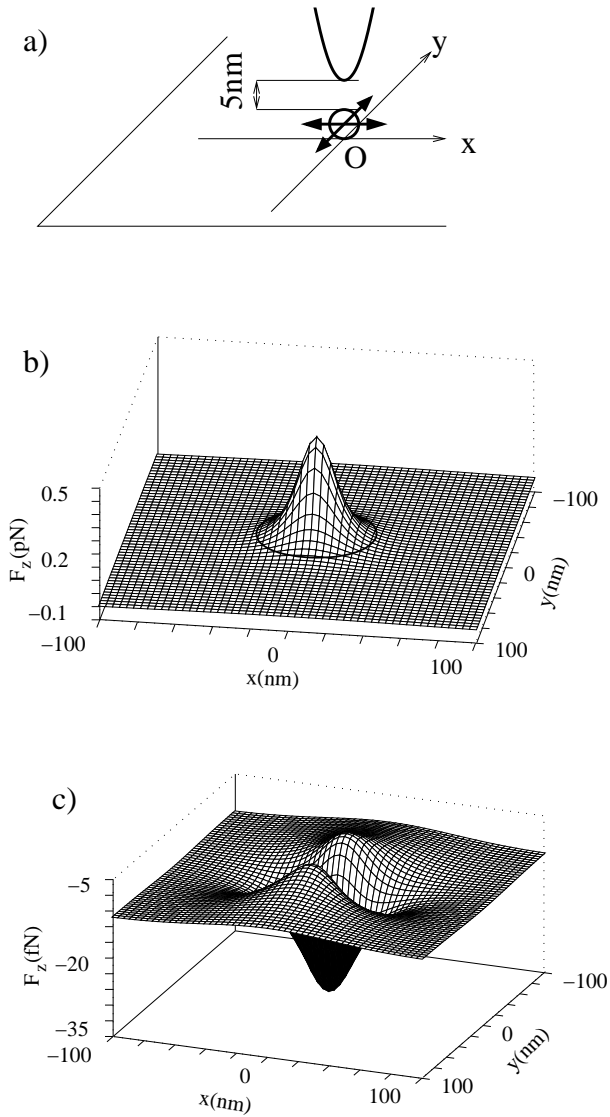}
\end{center}
\caption{$z$ component of the force versus the position of the sphere
($a=10$~nm) when the tip is located at the origin. a) sketch of the
configuration. b) TM polarization. The thick line represents the case
$F_z=0$. c) TE polarization.}
\end{figure}

We see that when the tip is far from the particle the force is
negative: the sphere does not feel the tip. As the tip gets closer,
the particle starts to experience a positive force along $z$. The
change of sign of the $z$ component of the force occurs when the
sphere is about 30~nm away laterally from the tip. Below this distance
the sphere is in the area of the enhancement of the field at the tip
apex and the gradient force changes sign, hence the sphere is
attracted toward the tip.  The region where $F_z=0$ is represented by
a solid closed curve in Fig.4b.  \begin{figure}[tb]
\begin{center}
\includegraphics*[draft=false,width=80mm]{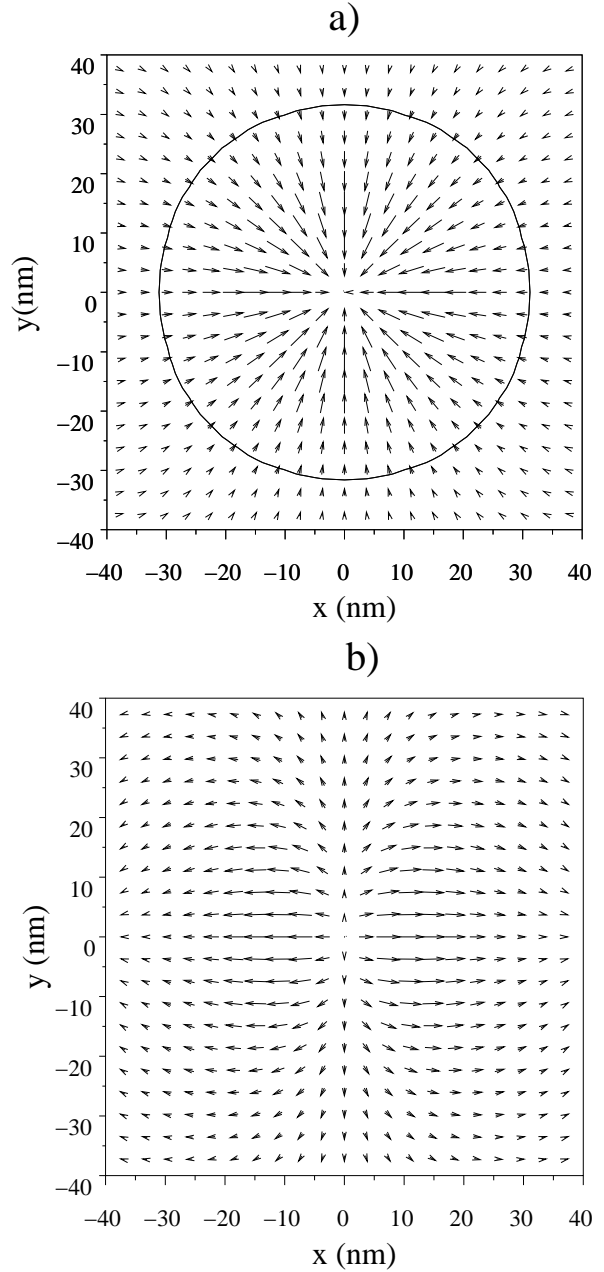}
\end{center}
\caption{Parallel component of the force versus the position of the
sphere when the tip is located at the origin. a) TM polarization. The
thick line represents the case $F_z=0$. b) TE polarization.}
\end{figure}If the tip is farther away from the
substrate, the zero force curve becomes smaller because of the
dependence of the force on the tip-surface distance (Fig. 2). If we
approximate the zero force curve by a circle, the radius of the circle
is about 7~nm when the tip is 31~nm above the substrate (and as shown
by Fig.2 it vanishes when the tip is 45~nm above the
substrate). Figure 4c, which pertains to TE polarization, shows that
the $z$ component of the force is always negative and smaller in
magnitude by a factor 100 than the force associated to the TM
polarization.  Note that the force becomes stronger (more negative)
when the sphere gets closer the tip.  Figure 5 represents the lateral
force ($\ve{F}_{\parallel}$) experienced by the sphere when the tip
scans the surface (the arrows represent the direction of the force
experienced by the sphere at the origin of the arrow, and the length
of the arrows shows the magnitude of this force). We only consider an
area of 40~nm around the origin as the lateral force decreases very
quickly away from the tip. In Fig. 5a, the vectors show that the
sphere is attracted by the tip, hence the lateral force pushes the
sphere toward the tip.  Therefore, when the tip and the particle are
close enough to each other for the $z$ component of the force to be
positive (the zero force, $F_z=0$ is always represented by the black
circle), and large enough to lift the particle off the surface, the
lateral force actually helps bringing the particle in the trap. This
effect is due to the symmetric illumination.  Again TE polarization
gives a different result. Figure 5b shows that the lateral force
pushes the particle away from the tip. However, since the magnitude of
the (downward) $z$ component of the force is larger than the $x$
component by a factor of 5, we expect that the sphere is not displaced
when the tip is scanned over it under TE illumination.  Note that
apertureless probes are often used in tapping mode when imaging a
surface. This mode minimizes the lateral motion imparted to the object
by the optical force.

We have shown that a tungsten probe can be used to trap efficiently a
nanometric object above a surface using TM illumination. For
nanomanipulation purposes it is important to assess the stability of the trap
as the probe lifts the particle off the substrate.  Figure~6 shows the
$z$ component of the force when the sphere is located at the apex of
the tip and the tip is moved vertically.  For the TM polarization
(Fig.~6a) the optical force remains positive over a large distance, at
least 200~nm.  The particle can therefore be manipulated vertically as
well as horizontally. The stability of the trap when the tip-particle
pair is away from the substrate prevents any disruptive interaction
with the surface roughness as shown in the next section. Note that the
evolution of the force versus the distance to the substrate is linear
rather than exponential. The particle experiences a negative gradient
force due the exponential decay of the intensity of the
illumination. At the same time, the particle experiences a positive
gradient force due to the field enhancement at the tip apex, which
also decreases exponentially with $z$ because this enhancement depends
on the intensity of the evanescent illuminating light.  The
competition between these two contributions results in a weak decrease
of the trapping force as the particle is moved away from the
substrate.

\begin{figure}[ntb]
\begin{center}
\includegraphics*[draft=false,width=80mm]{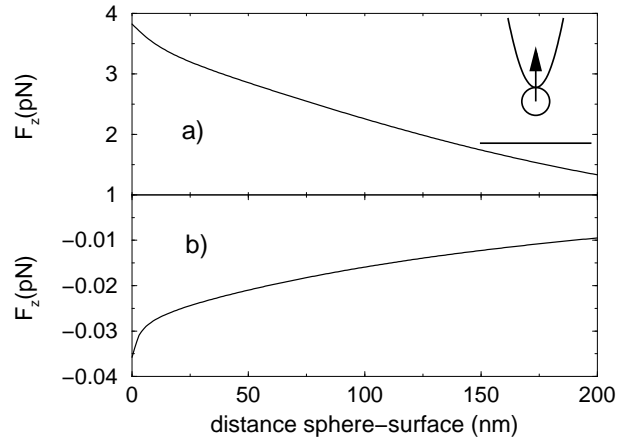}
\end{center}
\caption{$z$ component of the force experienced by the sphere as a
function of the distance between the sphere and the substrate. The
sphere is placed at the apex of the probe.  $\theta=43^{\circ}$.  a)
TM polarization.  b) TE polarization.}
\end{figure}

As described in \cite{prl}, the procedure to trap a small object with
a tungsten tip is the following: first TE illumination is
used while the tip scans the surface in tapping mode or in
constant-height mode if the area under investigation is small
enough. Such modes avoid the displacement of the particle by the tip.
Once an object has been selected, the probe is placed above the object
and the polarization of the illumination is rotated to TM. The probe
is then brought down over the particle and captures it. The probe can
then move the particle above the substrate, both horizontally and
vertically, to a new position (note that if, for some reason, one
wishes to move the particle over distances larger than the size of the
illumination spot, one could move \emph{the sample} with a
piezoelectric device once the sphere is trapped at the apex of the
tip).  As shown by Fig. 6b, as the $z$ component of the force in TE
polarization is always negative, the nanoparticle can be released by
switching back to TE polarization. The lack of trapping under TE
illumination is actually an important advantage during both the
imaging (selection) and release phases of the manipulation. Indeed,
under TE illumination, when the tip is above a particle, it actually
increases the downward optical force, which contributes to prevent the
tip from sweeping the particle away.

\subsubsection{Efficiency of the manipulation scheme}

In the previous section we have established the possibility to
manipulate selectively a nanoparticle above a flat dielectric
substrate. Now we study the influence of the different parameters
of the system (tip radius at the apex, angle of incidence, etc) on the
efficiency of the trap.

\paragraph{Influence of the illumination}

So far we have considered an angle of incidence of 43 degrees, which
corresponds to a slow decay of the evanescent field above the
substrate. \begin{figure}[ntb]
\begin{center}
\includegraphics*[draft=false,width=80mm]{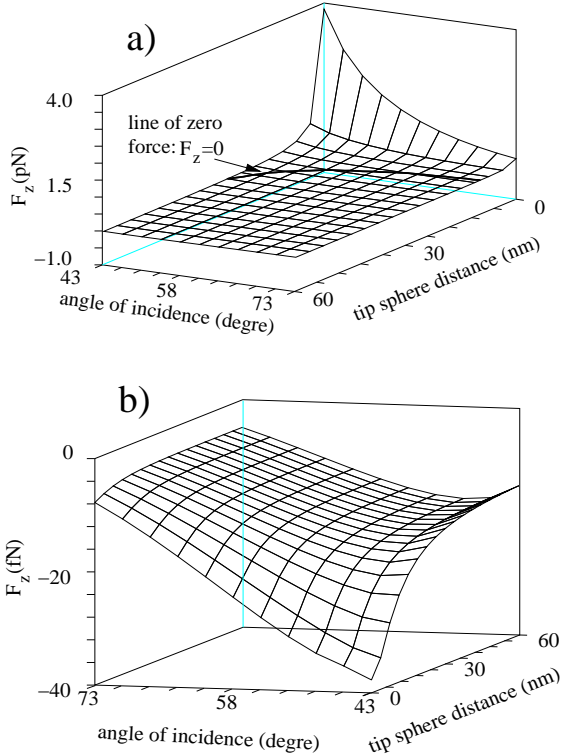}
\end{center}
\caption{$z$ component of the force experienced by the sphere as a
function of the distance between the tip and the sphere and the
substrate, and the angle of incidence.  a) TM polarization.  b) TE
polarization.}
\end{figure} 
\begin{figure}[ntb]
\begin{center}
\includegraphics*[draft=false,width=80mm]{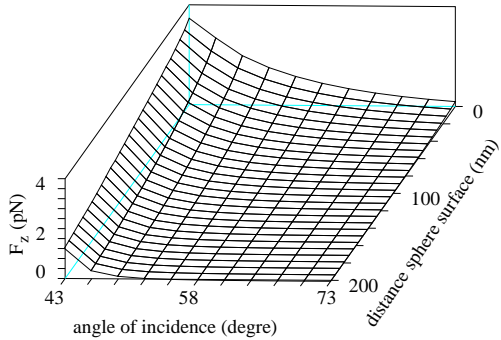}
\end{center}
\caption{$z$ component of the force experienced by the sphere as a
function of the distance between the sphere and the substrate with
$\theta=43^{\circ}$ for TM polarization.}
\end{figure}
Figures 7 and 8 show the influence of the angle of incidence on the
capability of the tip to manipulate the nano-object. Figure 7 shows
the evolution of the $z$ component of the force as the tip gets closer
the surface, versus the angle of incidence for both polarizations. In
Fig.7a (TM polarization) one can see that for a fixed distance between
the tip and the substrate, the larger the angle of incidence, the
smaller the magnitude of the force experienced by the trapped object.
Note that as the angle of incidence increases, the initial
(evanescent) field decays faster.  Accordingly, a weaker field reaches
the tip and the enhancement of the field at the tip apex is less
important leading to a smaller positive gradient force. As a
consequence, when the tip approaches the sphere, the change of sign
for the $z$ component of the force occurs for smaller tip-sphere
distances at larger angles of incidence (Cf. the thick line on Fig. 7a
which represents the level curve $F_z=0$).  For example for
$\theta=43^{\circ}$, $z$=25~nm and for $\theta=73^{\circ}$,
$z$=8~nm. This means that it is easier to manipulate the sphere when
the angle of incidence is close to the critical angle. Notice that for
TE polarization (Fig. 7b), when the angle of incidence increases, the
magnitude of the negative $z$ force decreases.  The explanation for
this evolution is similar to that of the TM polarization case: the
incident field that reaches the apex of the tip is weaker for large
angles of incidence. Accordingly, the interaction between the tip and
the particle becomes weaker as the angle increases, and the magnitude
of the repulsive force decreases. Figure 8 represents the $z$ force
experienced by the sphere when it is located at the apex of the tip,
versus the angle of incidence and the distance between the trapped
sphere and the substrate. The force along $z$ decays more rapidly for
larger angles of incidence. The exponential decay of the incident
field is stronger when the angle of incidence is far from the critical
angle. As the positive force along $z$ is due to the enhancement of
the field at the apex of the tip and it depends on the value of the
incident evanescent field at the tip apex, the $z$ component of the
force follows the same behavior as the incident field. Note that the
influence of the wavelength is easy to infer. The initial field decays
as $e^{-\gamma z}$ with
$\gamma=\sqrt{2\pi(\varepsilon\sin^2\theta-1)/\lambda}$, where
$\lambda$ is the wavelength in vacuum. When $\lambda$ increases,
$\gamma$ decreases, hence the exponential decay is slower and the
manipulation is easier to perform.

\paragraph{Influence of the geometry}

In the previous section, both the radius of the tip $r$, and the
radius of the sphere $a$ were 10~nm. In this paragraph we study the
influence of these two geometrical parameters. Figure 9 shows the
evolution of the force along $z$ versus $a$, for both polarizations,
and two angles of incidence. The tip ($r$=10~nm) is in contact with
the sphere. One might expect a force proportional to $a^3$ as the
gradient force is proportional to the real part of the polarizability,
hence to the volume of the sphere. Actually, this behavior is only
observed for the TE polarization (Fig. 9b).  For the TM polarization,
when $a$ increases, so does the distance between the tip and the
substrate.  Thus we have a competition between the increase of the
gradient force due to a larger $a$ and the decrease of the field
enhancement at the tip apex due to a larger distance between the tip
and the substrate. For the smallest angle ($\theta=43^{\circ}$),
because the decrease of the field is slow, the force starts by
increasing linearly for small values of $a$. When $a$ increases, the
enhancement of the field at the tip decreases and the positive
gradient force due to this enhancement vanishes.  The competition
between these two effects leads to a maximum of the force for
$a$=40~nm.  For the largest angle ($\theta=60^{\circ}$) the incident
field decays rapidly in that case, and for $a$ larger than 40~nm the
$z$ force experienced by the sphere becomes negative. Thus it would
not be possible to manipulate a larger particle. For TE polarization
the $z$ component of the force varies roughly as $a^3$. This implies
that the main contribution to the force is due to the incident field:
the tip has a very weak influence on the force experienced by the
sphere. To check this argument we plot the force along $z$ without the
presence of the tip (line with the ``+'' symbol in fig9.b). 
\begin{figure}[ntb]
\begin{center}
\includegraphics*[draft=false,width=80mm]{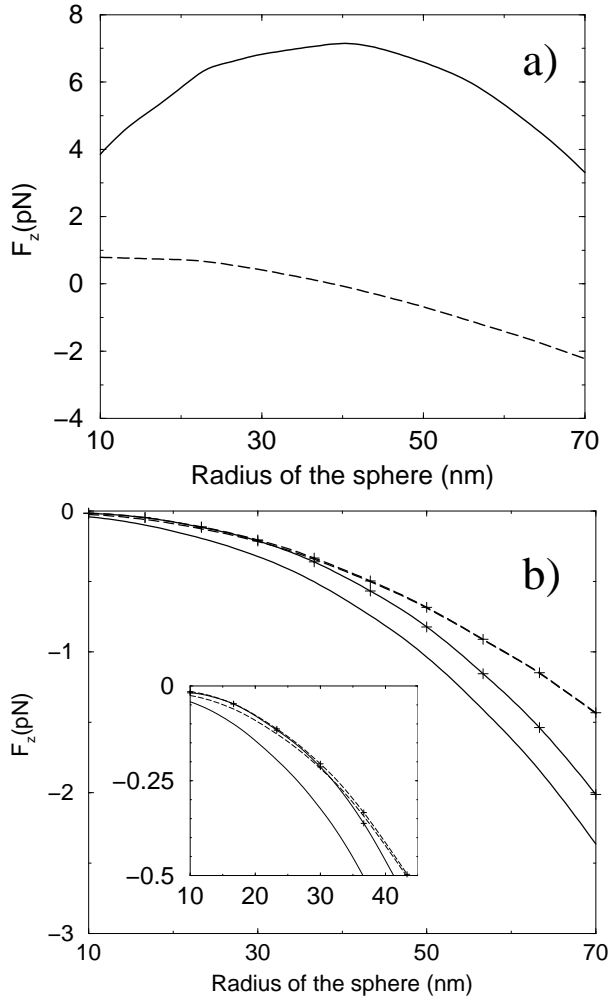}
\end{center}
\caption{$z$ component of the force experienced by the sphere as a
function of its radius. Plain line $\theta=43^{\circ}$, and dashed
line $\theta=60^{\circ}$ a) TM polarization.  b) TE polarization.  The
symbol + pertains to results for when the tip is not taken into
account in the computation.}
\end{figure}
These curves show that for the largest angle ($\theta=60^{\circ}$) the
above argument is true, only when the radius is small we can see a
slight shift of the force due to the presence of the tip (see inset in
fig.9b). For the smallest angle ($\theta=43^{\circ}$) the presence of
the tip shifts the force curves toward negative values, as explained
in the previous section. This holds even for a very large radius of
the sphere because of the slow evanescence of the incident field.

We must now check that for large radii it is always possible to
manipulate the sphere above the substrate. Figure 10 shows the $z$
component of the force at two different angles of incidence for two
different radii: $a=$30~nm and 50~nm. One can see that it is possible
to lift the spheres up to 200~nm above the surface without any
problem, even for the case $\theta=60^{\circ}$ with $a=30$~nm which
corresponds to a radius close to the limiting case (Fig. 9a shows
$F_z=0$ for $a=40$~nm if $\theta=60^{\circ}$). In that case, the force
is small and the trap is less robust than for smaller angles of
illumination. Therefore, the trapping scheme presented here works over
a wide range of particle sizes. Notice that although we could not
compute the largest radius that we could manipulate at
$\theta=43^{\circ}$ because it would require too many subunits, we can
estimate that spheres with a radius up to around 90~nm can be trapped
and manipulated.

\begin{figure}[ntb]
\begin{center}
\includegraphics*[draft=false,width=80mm]{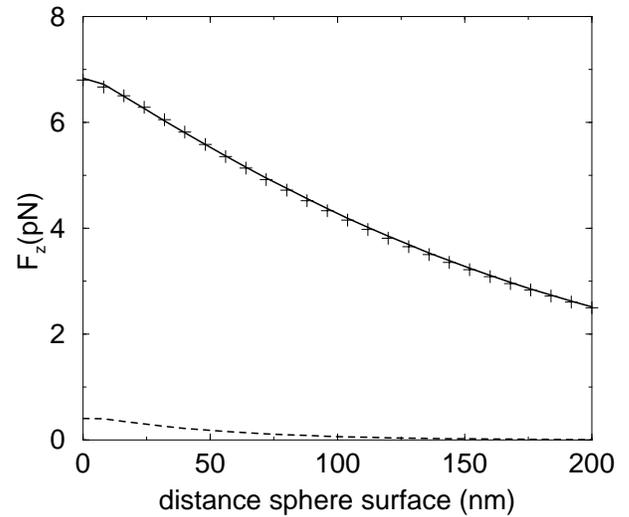}
\end{center}
\caption{$z$ component of the force experienced by the sphere in TM
polarization. Solid line: $a=$30~nm, $\theta=43^{\circ}$. Dashed line:
$a=$30~nm, $\theta=60^{\circ}$. Crosses: $a=$50~nm,
$\theta=43^{\circ}$.}
\end{figure}

Another relevant geometric parameter is the radius of curvature at the
apex of the tip. It is easy to see the importance of this parameter
for our optical trap because the enhancement of the field at the tip
apex depends directly on this radius. Figure 11 shows the $z$
component of the force versus $r$, for a particle with radius
$a=$10~nm, for two different angles of incidence. One can see that the
$z$ component of the force depends strongly on the radius of the tip
apex. For $\theta=43^{\circ}$ the squares are the CDM results and show
a decay of the force for larger radii.  The solid line is a fit of the
form $a_0/r+a_1$ where $a_0$ and $a_1$ are the parameters of the
fit. This form is associated to the $1/r$ dependence of the $z$ force,
which is found irrespective of the angle of incidence (see circles on
Fig. 11).

\begin{figure}[ntb]
\begin{center}
\includegraphics*[draft=false,width=80mm]{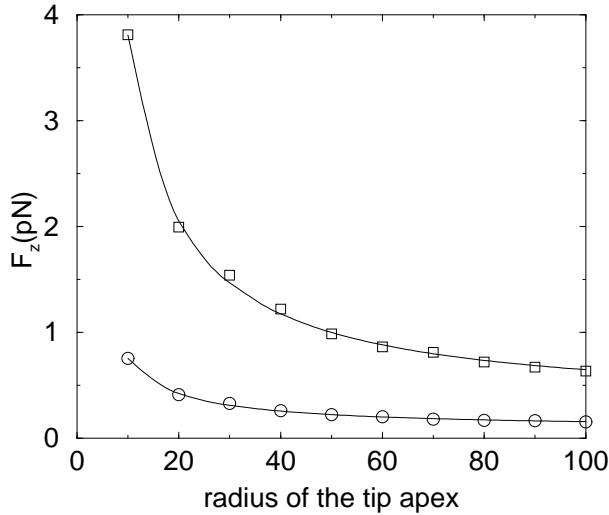}
\end{center}
\caption{$z$ component of the force experienced by the sphere for TM
polarization versus the radius of the tip apex. $a$=10~nm. Solid line
(fit of the form $a_0/r+a_1$) and squares (CDM results) for
$\theta=43^{\circ}$, and circles (CDM results) for
$\theta=60^{\circ}$}
\end{figure}

\subsection{Many particles on the surface}

In the preceding section we studied the case of an isolated sphere to
illustrate how to select and manipulate a nano-object above a
surface. It is nevertheless important to know whether the proposed
manipulation scheme would still work if several particles are
clustered together.  We consider a set of three spheres (radius 10~nm,
permittivity 2.25) aligned along the $x$ axis. The probe is placed
above the middle sphere. We account for the multiple scattering
between the three spheres, the substrate, and the tip. The optical
binding induced among the spheres~\cite{chaumet4} is also included in
our description. For this configuration again, TE illumination does
not permit trapping.  For TM illumination, we plot in Fig.~12 the $z$
component of the force experienced by the middle sphere and by those
on the sides as a function of the vertical distance between the probe
and the middle sphere.  For an angle of incidence $\theta=43^{\circ}$,
as the tip gets closer to the middle sphere, the $z$ component of the
force, although the strongest for the middle sphere, remains positive
for the two side spheres. This could be a problem if one wanted to
manipulate only one particle among several.  The central particle can
be selectively trapped by increasing the angle of incidence of the
illuminating beams to tighten the trap in the $x$ direction. In
Fig.~12 we see that for $\theta=60^{\circ}$ the optical force induced
by the probe is positive only for the middle sphere. This remains true
for three spheres aligned along $y$. Figure 13 shows the extraction of
the middle sphere by the tip.  Our calculation shows that the vertical
force experienced by the two side spheres remains negative when the
probe moves away from the substrate for the angle of incidence
$\theta=60^{\circ}$.  Therefore, the spheres on the sides do not
hinder the capture of the middle sphere if the angle of incidence is
adequately chosen.

\begin{figure}[ntb]
\begin{center}
\includegraphics*[draft=false,width=80mm]{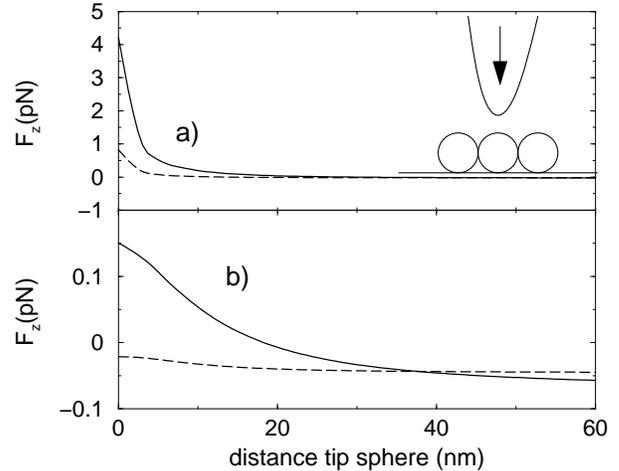}
\end{center}
\caption{$z$ component of the force experienced by the sphere in TM
polarization. Solid line $\theta=43^{\circ}$, dashed line
$\theta=60^{\circ}$. a) Force experienced by the middle sphere. b)
Force experienced by the side spheres.}
\end{figure} 

\begin{figure}[ntb]
\begin{center}
\includegraphics*[draft=false,width=80mm]{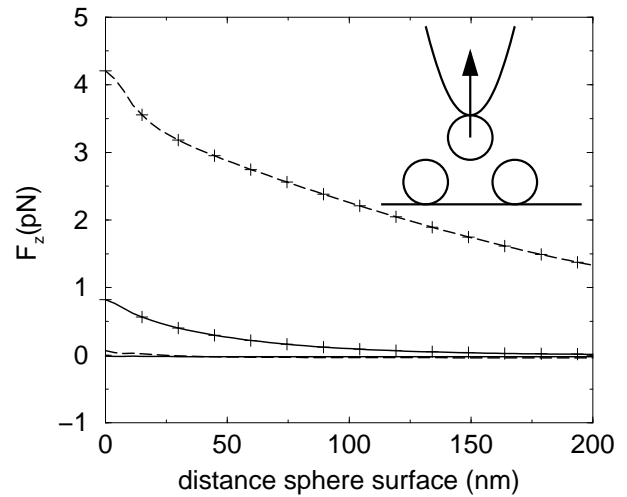}
\end{center}
\caption{$z$ component of the force experienced by the sphere in TM
polarization. Solid line $\theta=43^{\circ}$, dashed line
$\theta=60^{\circ}$. With symbol ``+'': force experienced by the
middle sphere. With no symbol: force experienced by the side spheres.}
\end{figure}

Notice that this computation is done for three identical spheres. We
show in Fig. 14 that if the middle sphere is larger than the other
two, it is still possible to trap the middle sphere without disturbing
the side spheres. Figure 14a shows the $z$ component of the force
experienced by the middle sphere for three different radii: $a$=10,
17, and 28~nm. The side spheres have a fixed radius: $a=$10~nm.  When
the middle sphere is lifted off the surface by the tip (angle of
incidence $\theta=60^{\circ}$) the force along $z$ is always positive,
hence the manipulation of the middle sphere is not disturbed by the
side spheres. Figure 14b shows the evolution of the $z$ force
experienced by the sides spheres during the extraction of the middle
sphere. This force is always negative, therefore the side spheres are
not attracted by the tip. Moreover, as the radius of the middle sphere
increases, the $z$ component of the force becomes larger while being
negative, thus excluding the possibility of having the side spheres
captured by the tip. This reflects the fact that as the radius of the
middle sphere increases, so does the distance between the side spheres
and the apex of the tip. Hence as the tip is farther from the sides
sphere its influence is weaker.

\begin{figure}[ntb]
\begin{center}
\includegraphics*[draft=false,width=80mm]{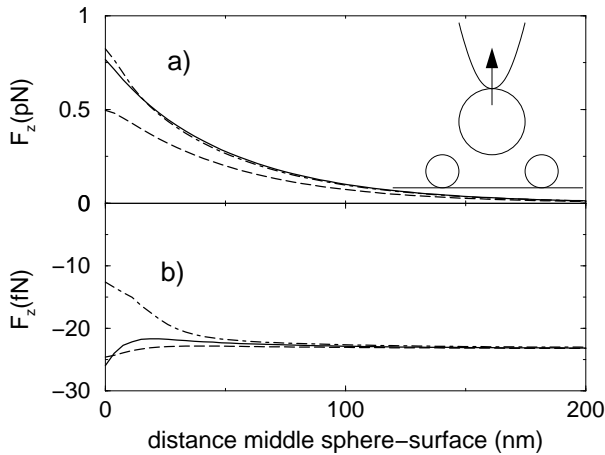}
\end{center}
\caption{$z$ component of the force experienced by the sphere in TM
polarization with $\theta=60^{\circ}$. The radius of the side spheres
is 10~nm. The radius of the middle sphere is: $a=10$~nm dot dashed
line, $a=$17~nm solid line, and $a=28$~nm dashed line a) Force
experienced by the middle sphere. b) Force experienced by the side
spheres.}
\end{figure}

In Fig.~15 we present the case where the middle sphere is smaller than
the side spheres.  First for an angle of incidence close to the
critical angle ($\theta=43^{\circ}$) the three spheres experience a
positive force when the tips approaches the substrate. If we look
carefully (see the inset), we see that the $z$ component of the force
becomes positive for the side spheres before the middle sphere
experiences a positive force. Hence it is impossible to only
manipulate the middle sphere. If the angle of incidence is increased
to the value $\theta=60^{\circ}$ to make the trap smaller, we show
that the $z$ component of the force becomes positive for the middle
sphere first.  However, when the tip is in contact with the middle
sphere the $z$ force is positive for all three spheres. Notice that to
get a smaller trap one can increase the angle of incidence but in that
case the force would be very small compare to the other force in the
system (see appendix A and B). Therefore, in this case it is
impossible to capture selectively the middle sphere. The solution
would be to first move the side spheres aways to isolate the smaller
middle sphere, and only after would it be possible to trap it. Notice
that when the tip is far from the surface the negative force is
stronger for the largest spheres. This is due to the gradient force
proportional to $a^3$.

\begin{figure}[ntb]
\begin{center}
\includegraphics*[draft=false,width=80mm]{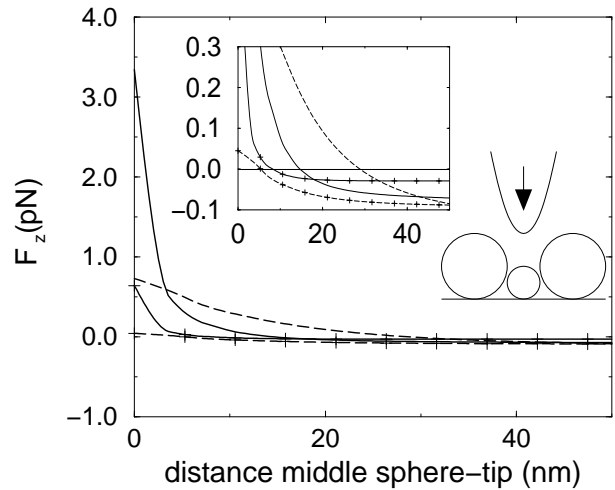}
\end{center}
\caption{$z$ component of the force experienced by the sphere in TM
polarization The radius of the middle sphere is 10~nm. The radius of
the side sphere is $a=$17~nm.  Solid line: force experienced by the
middle sphere. Dashed line: force experienced by the side
spheres. Curve with no symbol: $\theta=43^{\circ}$. Curve with symbol
``+'': $\theta=60^{\circ}$. The inset is a magnification of what
happens close to the sign reversal for the force along $z$.}
\end{figure}

In the previous section we showed that a tungsten probe can be used to
trap efficiently a nanometric object above a surface using TM
illumination. By moving the tip laterally, it is possible to transport
the selected particle in a precise manner. However, we must check that
the electromagnetic field scattered by another particle on the
substrate would not disturb the trap during the transport.  Figure 16
shows the evolution of the force experienced by the sphere trapped by
the tip when a second sphere is on the surface as show by Fig.~16a
(both spheres have a radius of 10~nm). Figure 16b shows that the $z$
component of the force on the trapped sphere is not altered by the
presence of the other sphere. When the two spheres are 30~nm apart
($h=50$~nm), the force along $z$ does not depend on the position of
the tip. Figure 16c shows that the lateral force is more sensitive and
we can see some oscillations when the two spheres are close to each
other. However, this is not really a problem since even if the two
spheres are only 30~nm apart, the lateral force is a thousand times
smaller that the force along $z$, and therefore, would not hinder the
optical trapping. We have studied the case where the tip and the
trapped sphere scan the surface at $h=20$~nm. When the tip is located
at the origin, the two spheres are in contact (the trapped sphere and
the sphere on the substrate) and the $z$ component of the force
becomes negative, of the order of -5~pN, and the lateral force when
the spheres are close to each other is about 1~pN.  In this
configuration the trapped sphere can escape and thus be lost by the
tip. In summary, if the distance between the two spheres is larger
than three times the radius of the sphere on the surface, then the
trapped sphere is not disturbed.

\begin{figure}[ntb]
\begin{center}
\includegraphics*[draft=false,width=80mm]{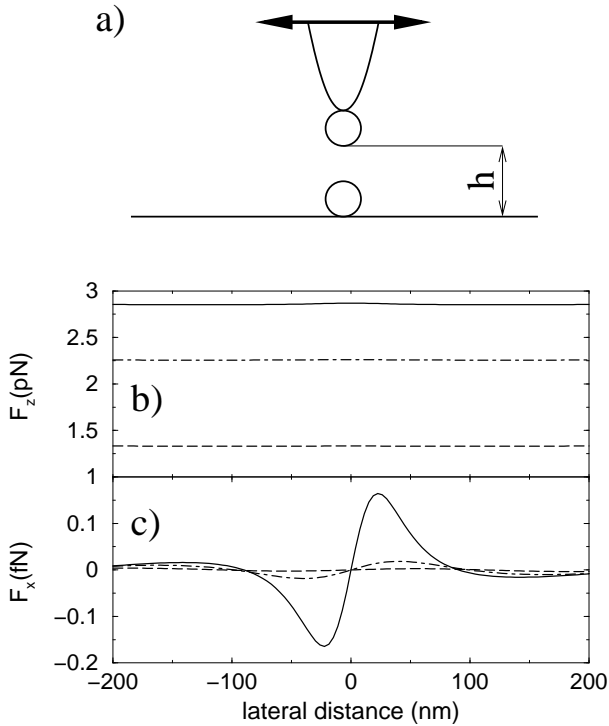}
\end{center}
\caption{Force experienced by a sphere ($a=$10~nm) trapped at the apex
of the tungsten tip, when the tip scans the substrate at different
height $h$, with another particle ($a=10$~nm) on the surface localized
at the origin.  Solid line: $h=50$~nm. Dot-dashed line:
$h=100$~nm. Dashed line: $h=200$~nm. a) sketch of the
configuration. b) $z$ component of the force. c) $x$ component of the
force.}
\end{figure}

\section{conclusion}

We have presented a detailed study of a trapping scheme that allows
one to trap and nanomanipulate, in a selective manner, nanometric
particles in air above a substrate.  The substrate is illuminated
under total internal reflection by two laser beams which create two
counter-propagating evanescent waves. An apertureless tungsten probe
is used to scatter these two waves and generate a localized optical
trap.  An object of a few nanometers can be selectively brought into
the trap and manipulated with the probe.  An important advantage of
this scheme is the possibility to use the probe to localize the
particles upon the surface. Using TE polarization the tip can scan the
surface in tapping mode or constant-height mode, and allow one to
acquire an optical near-field image of the surface.  Because in TE
polarization the $z$ component of the optical force is directed toward
the substrate, there is no risk of displacing the particles during the
imaging phase.  Then, just by switching to TM polarization, we can
manipulate the particles.  As we showed, even if many particles are
clustered, varying the angle of incidence still makes it possible to
manipulate selectively only one particle.

An interesting extension of this work will be a study of the influence
of different illuminations (e.g. focused beam), and the study when the
particle is either absorbing or metallic. In that case the optical
force has two contributions: the gradient force and the momentum
transfer from the laser to the particle due to absorption. For
metallic particles, the strong spectral dependence of the
electromagnetic response (or the resonances in the response of
dielectric and metallic particles) could lead to new effects. For
example at some wavelength the gradient force on a silver particle
vanishes, and only the absorbing force remains. Such phenomena could
lead to a material selective trapping.  It will also be interesting to
explore the possibility of trapping a small gold particle, a few
nanometers in size, and use it as a highly localized probe for
topographic or spectroscopic studies \cite{sqalli,kalkbrenner}.

\appendix

\section{Importance of the evanescent illumination \label{A1}}

In this appendix we show that the choice of total internal reflection
illumination is the most adequate to get a strong optical
force. Figure 17 shows the $z$ component of the force when the angle
of incidence is varied between 0 and 90 degrees for an illumination
either from above or from below the surface. For an illumination from
below the surface (internal reflection), and for TM polarization, the
largest force is obtained for $\theta=\theta_c$.  The magnitude of the
force decreases exponentially when $\theta$ increases. For $\theta$
smaller than the critical angle, the force is small and even
negligible for $\theta$ smaller than 20 degrees. For TE polarization
the minimum of the $z$ component of the force is obtained for
$\theta=43^{\circ}>\theta_c$. This negative force is important as it
prevents the tip from displacing the particle during scans. If the
surface is illuminated from above, for TM polarization the $z$
components of the force remains weaker that the force obtained with an
evanescent wave, and for TE polarization the negative force is very
weak compared to those obtained with an evanescent wave above the
surface.  Hence to trap and manipulate a nano-object it is best to
choose an angle of incidence close to but larger than the critical
angle.

\begin{figure}[ntb]
\begin{center}
\includegraphics*[draft=false,width=80mm]{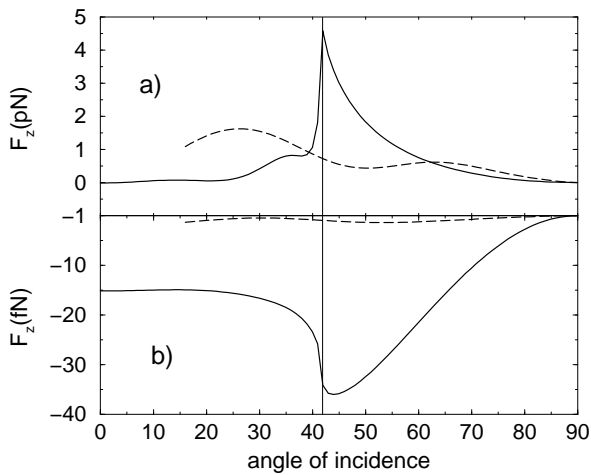}
\end{center}
\caption{$z$ component of the force experienced by the sphere versus
the angle of incidence.  The vertical line is plotted for
$\theta=\theta_c$ in the case of total internal reflection. The solid
line pertains to the substrate illuminated from below the surface, and
the dashed line pertains to an illumination from above the surface. a)
TM polarization. b) TE polarization.}
\end{figure}

\section{Discussion on the role of forces other than the optical
force\label{A2}}

In an actual experiment, there would be other forces attracting the
sphere toward the substrate. These forces are mainly four: van der
Waals, electrostatic, capillary, and gravitational forces. One must
therefore compare their effect with that of the optical forces in
order to assess the robustness of the scheme.

\subsection{The van der Waals force}

The van der Waals force~\cite{feiler} between two particles (which is
the Casimir force in the nonretarded case~\cite{casimir}) can be
described as a short range force, derived from the Lennard-Jones 6-12
potential, in the form \be
F_w=\frac{AS}{6h^2}\left(\frac{z^6}{4h^6}-1\right), \ee $h$ being the
distance between the two particles, $A$ the Hamaker constant
($A=60$~zJ for glass, and about 200~zJ for tungsten), $z$ corresponds
to the separation of lowest energy between two particles (i.e. the
position of the minimum of the Lennard-Jones potential) that we have
estimated at 0.5~nm, and $S$ is the Derjaguin geometrical factor
related to the mutual curvature of the two particles. Notice that the
van der Waals force is maximum when $z=h$.

If we compute the van der Waals force between the glass sphere and the
glass surface we have $S=a$ ($a$ radius of the sphere), hence the
force is $F_w=0.3$~pN. This force is not a problem as the optical
force when the tip is in contact with the sphere is larger than
1~pN. In addition we have to consider the van der Waals force between
the apex of the tip and the sphere when they are in contact.  We have
estimated it for the case where the radius of the sphere and the
curvature of the tip apex are equal to 10~nm, with $S=a/2=5$~nm,
$A=\sqrt{60 \times 200}$~zJ, for which we obtain $F_w=0.27$~pN. Hence
the two van der Waals forces are of comparable magnitude and cancel
each other out when the particle is in contact with both the substrate
and the tip. Therefore the van der Waals force does not hamper the
manipulation of the particle. A problem can arise when we want to
release the sphere from the trap, as switching back to TE polarization
may not create a strong enough repulsive force.  There are many ways
to avoid this problem. One is to approach the tip-sphere system to the
surface to benefit from the van der Waals force between the substrate
and the sphere. Another solution is to choose an angle of incidence
close to the critical angle to increase the magnitude of the repulsive
force.  Of course, one could also increase the intensity of the
incident field.

Notice that the van der Waals force is computed here for perfectly
smooth bodies. In reality this force should be weaker. Indeed, for a
surface roughness about 2~nm, the van der Waals force is reduced by a
factor of 10.~\cite{arai}

\subsection{The electrostatic force}

The electrostatic force (Coulomb force) between an electrically
charged sphere and an uncharged plane can be expressed as~\cite{arai}
\be F_e=\frac{\pi}{\varepsilon_0}
\frac{\varepsilon-1}{\varepsilon+1}a^2\sigma^2,\ee $\sigma$ being the
charge surface density ($10^{-3}$~Cm$^{-2}$ in very dry conditions),
$\varepsilon_0=8.85.10^{-12}$~Fm$^{-1}$ the permittivity of vacuum,
and $\varepsilon$ the relative permittivity of the dielectric
substrate. For $a=10$~nm and $\varepsilon=2.25$, we get
$F_e=0.013$~pN. This force is clearly negligible compared to the
optical force. And moreover this force will be weaker for a
conductor.~\cite{arai}

\subsection{The capillary force}

If there is water on the surface, there will be a capillary force
which can be expressed as~\cite{arai} \be F_c=2\pi a \gamma,\ee
$\gamma$ being the surface tension of water. With
$\gamma=72.10^{-3}$~Nm$^{-1}$. For a radius of the sphere $a=10$~nm we
get $F_c=4.5$~pN. This force is of the same order as the optical force
hence it is necessary to work in a dry environment in order to
reduce this capillary force.

\subsection{The gravitational force}

The force of gravity is \be F_g=mg=\frac{4}{3} \pi a^3 \rho g \ee
where $g=10$~ms$^{-2}$ is the gravitational acceleration and
$\rho=2500$~kgm$^{-3}$ is the density of glass.  If the radius is
equal to 10~nm we find $F_g=0.1$~aN, and the $z$ component of the
optical force experienced by the sphere is larger (by a factor $10^7$)
than the gravitational force. Hence gravity can be neglected.

\subsection{Conclusion}

In conclusion, in a dry environment (no capillary force) only the van
der Waals force could perturb the release of the particle but as we
mentioned previously, this force becomes weaker when roughness is
taken into account. Note, however, that these four forces do not
depend of the illumination whereas the optical forces depends of the
intensity of the incident field. Therefore, one solution to avoid any
disruptive contribution from the van der Waals force is to increase
the power of the laser beam. For example, with the power used by
Okamoto and Kawata~\cite{okamoto}, which corresponds to an irradiance
of 0.2~W/$\mu$m$^2$, the optical force is multiplied by a factor 4
compared to the computation presented in this manuscript. Another way
of increasing the optical force is to choose another material for the
probe. For example, at a wavelength of 450~nm, a silver tip when in
contact with the sphere, generates an optical force six times stronger
than that created by a tungsten tip at $\lambda=514.5$~nm.

%%%%%%%%%%%%%%%%%%%%%%%%%%%REFERENCES%%%%%%%%%%%%%%%%%%%%%%%%%%%%

\end{multicols}


\begin{references}


\bibitem{ashkin69}\label{ashkin69} A. Ashkin, Phys, Rev. Lett. {\bf
24}, 156 (1970).

\bibitem{ashkin70}\label{ashkin70} A. Ashkin, Phys. Rev. Lett. {\bf
25}, 1321 (1970).

\bibitem{ashkin86} A. Ashkin, J. M. Dziedzic, J. E. Bjorkholm, and
S. Chu, Opt. Lett. {\bf 11}, 288 (1986).

\bibitem{ashkin87} A. Ashkin, J. M. Dziedzic, and T. Yamane, Nature
{\bf 330}, 769 (1987).

\bibitem{block89} S. M. Block, D. F. Blair, and H. C. Berg, Nature
{\bf 338}, 514 (1989).

\bibitem{ashkin97} A. Ashkin, Proc. Natl. Acad. Sci. USA {\bf 94},
4853 (1997).

\bibitem{burns} M. Burns, J.-M Fournier, and J. Golovchenko,
Phys. Rev. Lett. {\bf 63}, 1233 (1989).

\bibitem{landragin} A. Landragin, J.-Y. Courtois, G. La\-bey\-rie,
N. Van\-steen\-kiste, C. I. Westbrook, and A. Aspect,
Phys. Rev. Lett. {\bf 77}, 1464 (1996).

\bibitem{assembling} R. Holmlin, M. Schiavoni, C. Chen, S. Smith,
M. Prentiss, and G. Whitesides, Angew. Chem. Int. Ed. {\bf 39}, 3503,
(2000); E. R. Dufresne, G. C. Spalding, M. T. Dearing, S. A. Sheets,
and D. G. Grier, Rev. Sci. Inst. {\bf 72}, 1810 (2001). E.
R. Dufresne and David G. Grier, Rev. Sci. Instr. {\bf 69}, 1974
(1998).

\bibitem{macdonald}\label{macdonald} M. P. Macdonald, L. Paterson,
K. Volke-Sepulveda, J. Arlt, W. Sibbet, K. Dholakia, Science {\bf
296}, 1101 (2002).

\bibitem{gustavson} T. L. Gustavson, A. P. Chikkatur, A. E. Leanhardt,
A. G{\"o}rlitz, S. Gupta, D. E. Pritchard, and W. Ketterle,
Phys. Rev. Lett. {\bf 88}, 020401 (2002).

\bibitem{stm} S. Hla, L. Bartels, G. Meyer, and K. Rieder,
Phys. Rev. Lett. {\bf 85}, 2777 (2000); T. W. Fishlock, A. Oral,
R. G. Edgell, and J. B.  Pethica, Nature {\bf 404}, 743 (2000);
H. C. Manoharan, C. P. Lutz, and D. M. Eigler, Nature, {\bf 403}, 512
(2000).

\bibitem{prl} P. C. Chaumet, A. Rahmani, and M. Nieto-vesperinas,
Phys. Rev. Lett. {\bf 88}, 123601 (2002). 

\bibitem{zenhausern} F. Zenhausern, Y. Martin, and
H. K. Wickramasinghe, Science {\bf 269}, 1083 (1995); R. Bachelot,
P. Gleyzes, and A. C. Boccara, Opt. Lett. {\bf 20}, 1924 (1995).

\bibitem{defornel} F. de Fornel, {\it Evanescent Waves}, Springer
series in Optical Sciences, vol.73 (Springer Verlag, Berlin, 2001).

\bibitem{chaumet1} P. C. Chaumet, and M. Nieto-Vesperinas,
Phys. Rev. B. {\bf 61}, 14119 (2000); {\bf 62}, 11185 (2000)

\bibitem{purcell}\label{purcell} Purcell, and Pennypacker,
Astrophys. J. {\bf 186}, 705 (1973).

\bibitem{chaumet3} P. C. Chaumet, A. Rahmani, F. de Fornel, and J.-P
Dufour, Phys. Rev B {\bf 58}, 2310 (1998).
                                    
\bibitem{draine} B. T. Draine, Astrophys. J. {\bf 333}, 848 (1988).

\bibitem{jackson75} J. D. Jackson, {\it Classical Electrodynamics},
2nd ed. (John Wiley, New York, 1975), p395.

\bibitem{rahmani1} A. Rahmani and G. W. Bryant, Opt. Lett. {\bf 25},
433 (2000).

\bibitem{agarwal} \label{agarwal} G. S. Agarwal, Phys. Rev. A {\bf
11}, 230 (1975); {\bf 12}, 1475 (1975).


\bibitem{rahmani2} A. Rahmani, P. C. Chaumet, and F. de Fornel,
Phys. Rev A {\bf 63}, 023819 (2001).

\bibitem{time} In fact the optical force is the time averaged force as
shown in Ref.[\ref{chaumet2}].

\bibitem{chaumet2}\label{chaumet2} P. C. Chaumet, and
M. Nieto-Vesperinas, Opt. Lett.  {\bf 25}, 1065 (2000).

\bibitem{novotny}\label{novotny} L. Novotny, R. X. Bian, and X. Sunney
Xie, Phys. Rev. Lett. {\bf 79}, 645 (1997).

\bibitem{chaumet4}\label{chaumet4} P. C. Chaumet, and
M. Nieto-Vesperinas, Phys. Rev. B. {\bf 64}, 035422 (2001).

\bibitem{kawata} S. Kawata, and T. Sugiura, Opt. Lett. {\bf 17}, 772
(1992).

\bibitem{sqalli} O. Sqalli, M.P. Bernal, P. Hoffmann, and
F. Marquis-Weible, Appl. Phys. Lett. {\bf 76}, 2134 (2000).

\bibitem{kalkbrenner} T. Kalkbrenner, M. Ramstein, J. Mlynek, and
V. Sandoghdar, J. Microsc. {\bf 202}, 72 (2001).

\bibitem{feiler} A. Feiler, I. Larson, P. Jenkins, and P. Attard,
Langmuir, {\bf 16}, 10269 (2000).

\bibitem{casimir} H. B. G. Casmir, and D. Polder, Phys. Rev. {\bf 73},
360 (1948).

\bibitem{arai} F. Arai, IEEE/RSJ Conf. on Intell. Robots and systems,
vol 2:236, Pittsburgh, PA, 1995.

\bibitem{okamoto} K. Okamoto and S. Kawata, Phys. Rev. Lett. {\bf 83},
4534 (1999).

\end{references}
\end{document}